\documentclass[acus]{JAC2003}

\usepackage[english]{babel}
\usepackage{graphicx}
\usepackage{amssymb}

\begin{document}
\title{SUPERCONDUCTIVITY ABOVE $H_{c2}$ AS A PROBE FOR NIOBIUM RF-CAVITY SURFACES}

\author{Sara Casalbuoni\thanks{scasalbu@physnet.uni-hamburg.de}, Lars von Sawilski, and J\"urgen K\"otzler\\
Institut f\"ur Angewandte Physik und Zentrum f\"ur Mikrostrukturforschung, Universit\"at Hamburg\\ 
Jungiusstrasse 11, D-20355 Hamburg, Germany}

\maketitle

\begin{abstract}
From the magnetic response of Nb-cylinders to time-varying fields the AC-conductivity, $\sigma'-i\sigma''$, and the critical current density $J_c^s$ are determined above the upper critical field $H_{c2}$.
The field dependencies of $\sigma'$ and $\sigma''$ allow us to identify the nucleation of incoherent surface superconductivity (SSC) at $H_{c3}$ and of coherent SSC at $H_{c3}^c\approx 0.81H_{c3}$.
The latter result turns out to be independent on different surface roughnesses and surface impurities obtained by surface treatments like chemical (BCP) and electrolytical polishing (EP), and low temperature baking (LTB).
They all have a large impact on $H_{c3}$ itself, which is associated with a change of electron mean free path $\ell$ due to impurities in a surface layer.
A detailed analysis of $\sigma'$, $\sigma''$ and  $J_c^s$ near $H_{c3}^c$ reveals that the coherence of the SSC results from a percolation transition.
For EP cylinders, this transition appears to be strictly two dimensional (2D), while the analysis for BCP cylinders with a rougher surface reveal much smaller $J_c^s$  and indicate a crossover to 3D.
As a rather surprising feature, we detected large concentrations of paramagnetic moments, which increased under LTB and were reduced by EP.\end{abstract}

\section{INTRODUCTION}

By applying the Ginzburg-Landau (GL) equations to an ideal surface 
adjacent to a vacuum interface in a magnetic field parallel to the surface, St.~James and De~Gennes~\cite{Stjames} have shown that superconductivity persists up to $H_{c3}~=~1.695H_{c2}$ in a surface sheath of thickness of the order of the coherence length $\xi$.
Hence SSC probes the surface in a layer of the order of $\xi$, which for Nb is~$\sim~50~\mbox{nm}$.
Microwave fields of~$\sim 1~\mbox{GHz}$ applied to operate Nb cavities have a penetration depth $\lambda \sim ~50~\mbox{nm}$.
In order to explore the effects of the different surface treatments applied for cavities production
on this interesting layer, we investigate in this contribution several features of the SSC.

\section{EXPERIMENTAL}

From Nb sheaths provided by W. C. Heraeus~\cite{heraeus} for rf cavities production raw cylinders
(2.5 $\times$ 2.8 mm$^2$) have been electroeroded.
Since during electroerosion a surface layer of
several $\mu\mbox{m}$ thick of Nb$_2$O$_5$ is formed the samples have been
afterwards about 50 $\mu\mbox{m}$ chemical etched (BCP) in a 1:1:2 mixture of HNO$_3$
(65$\%$), HF ($40\%$) and H$_3$PO$_4$ ($85\%$) at room temperature. 
To remove residual acids water rinsing has been
then applied. The next step has been annealing of about 2 hours at 800
$^{\circ}$C ($p < 10^{-7}$~mbar), which removes dissolved
hydrogen and relieves mechanical stress in the
material. To reproduce the standard
procedure of cavity treatment final chemical etching of about 50~$\mu$m 
and water rinsing has been performed (cylinder C).
In order to study the effects of electropolishing (EP) the lateral surfaces of some of the cylinders has been electropolished about $80~\mu\mbox{m}$ at room temperature in a mixture of HF ($56\%$) and H$_2$SO$_4$ ($90\%$) (cyl. E).
The effects of low temperature baking (LTB) have been studied by annealing a BCP (cyl. Cb) and an EP (cyl. Eb) sample, prepared as described above, for 48 hours at $120~^{\circ}\mbox{C}$ and $p <10^{-7}$~mbar.
On one of the BCP baked samples chemical etching $10\mu$m and water rinsing has been further applied (cyl. Cbe), in order to investigate whether eventual baking effects are lost.
Scanning Electron Microscopy (SEM) and Atomic Force Microscopy (AFM) analysis~\cite{gil}
have shown that the surface roughness of $\approx 1~\mbox{nm}$ measured on the grains on areas of about $10\times 10~\mu\mbox{m}^2$ is unchanged by EP and LTB, while steps at the grain boundaries, present in the BCP samples, are removed by EP and left unchanged by LTB.

The magnetic moments $m$ have been measured by using a commercial SQUID magnetometer (Quantum Design MPMS$_2$) at temperatures ranging from 2~K to 300~K in external fields up to 10~kOe.
The SQUID magnetometer allowed also to measure the AC susceptibility at 10~Hz.
Still using the so-called mutual inductance technique, another magnetometer has been used together with a conventional lock-in amplifier (EG\&G PARC Model 5302) to extend the AC susceptibility measurements up to frequencies of 1~MHz and in the temperature range between 1.5~K and 4.2~K (He4-pumped cryostat). 
The linearity of the response has been checked by varying the excitation amplitudes from 0.1~mOe to 5~Oe.
In all measurements the external magnetic fields (DC and AC) were aligned parallel to the long axis of the cylinders.
The demagnetization factor $N_Z=0.36$ resulting from the initial slope of the DC-magnetization $M=-H/(1-N_Z)$ agreed with the theoretical expression
$N_Z=1-1/(1+qa/b)$~ \cite{Brandt2}.
Here $q=4/3\pi+2/3\pi$~tanh$(1.27b/a$~ln$(1+a/b))$, where $b/a$ is the ratio between the thickness and the diameter of the cylinder.

\section{BULK SUPERCONDUCTIVITY}
\begin{figure}[htb]
\centering
\includegraphics*[width=65mm]{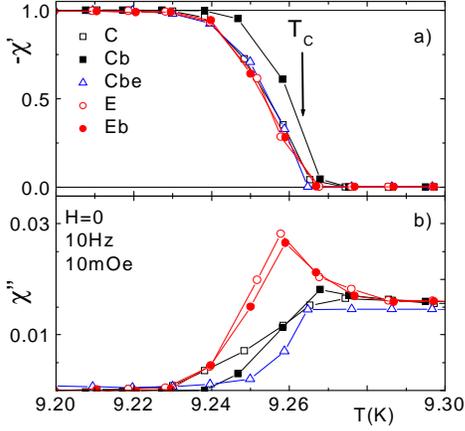}
\caption{a) Real and b) imaginary parts of the linear AC-susceptibility recorded near the zero-field transition temperatures of the Nb-cylinders under investigation. Note the different $\chi''$-scale in b).}
\label{Tc}
\end{figure}

We have determined the superconducting transition temperature $T_c$ from the onset of the screening component 
of the linear AC susceptibility  measured in zero field as a function of temperature (see Fig.\ref{Tc}).
The critical temperature $T_c=9.263(3)~\mbox{K}$ of all samples agrees with $T_c=9.25(1)~\mbox{K}$ reported by Finnemore at al.~\cite{finne} for high purity Nb ($RRR=1600(400)$).

From the linear AC susceptibility $\chi(\omega)$ in the normal conducting regime we have also evaluated the electrical resistivity  $\rho_n(\omega)$ using the method described in~\cite{Lars}.  
All samples exhibit an Ohmic resistivity $\rho_n(10~\mbox{Hz}, T\gtrsim T_c)=0.05(1)~\mu \Omega~\mbox{cm}$, confirming the purity of the samples and the specifications of the manufacturer of $RRR\simeq 300$.

\begin{figure}[htb]
\centering
\includegraphics*[width=65mm]{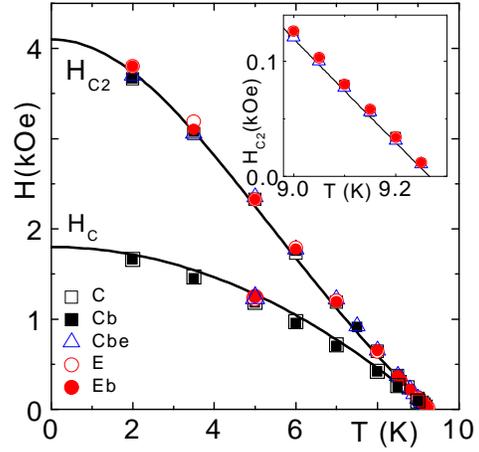}
\caption{Temperature variation of the upper critical field $H_{c2}$ and of the thermodynamical critical field $H_c$ of all cylinders, obtained as described in Ref.~\cite{Lars}, fitted to the phenomenological Ginzburg-Landau (GL)-type (solid curves) laws. The expected linear behaviour of $H_{c2}$ close to $T_c$ is displayed in the inset.}
\label{Hc2}
\end{figure}

 Following the procedure explained in Ref.~\cite{Lars}, we have determined the upper $H_{c2}$ and the thermodynamical critical field $H_c$ from the isothermal magnetization loops. Their temperature variation is indicated in Fig.~\ref{Hc2}.
The fit to the empirical law $H_c(T)=H_c(0)(1-(T/T_c)^2)$, gives for all samples $H_c(0)=1.80(5)~\mbox{kOe}$, which is  smaller than 1.99(1)~kOe determined in Ref.~\cite{finne}.
The values of the upper critical field have been fitted to the empirical temperature dependence 
\begin{equation}
H_{c2}(T)=H_{c2}(0)\frac{(1-(T/T_c)^2)}{(1+(T/T_c)^2)},
\label{eq:Hc2}
\end{equation}
yielding $H_{c2}(0)=4.10(5)~\mbox{kOe}$ in agreement with the literature values~\cite{finne}.
Within the GL theory it is then possible to determine for all samples the GL parameter $\kappa(0)=H_{c2}(0)/(\sqrt{2} H_{c}(0))=1.61(7)$ ~\cite{tinkham}, the GL coherence length $\xi(0)=\sqrt{\Phi_0/2\pi\mu_0 H_{c2}(0)}=28.3(2)~\mbox{nm}$ 
where $\Phi_0=2.07\times 10^{-15}~\mbox{Vs}$ and the GL penetration depth $\lambda(0)=\kappa(0) \xi(0)=46(2)~\mbox{nm}$.

As expected, LTB and/or EP do not change the bulk properties.

\section{SURFACE SUPERCONDUCTIVITY}

\begin{figure}[ht]
\centering
\includegraphics*[width=65mm]{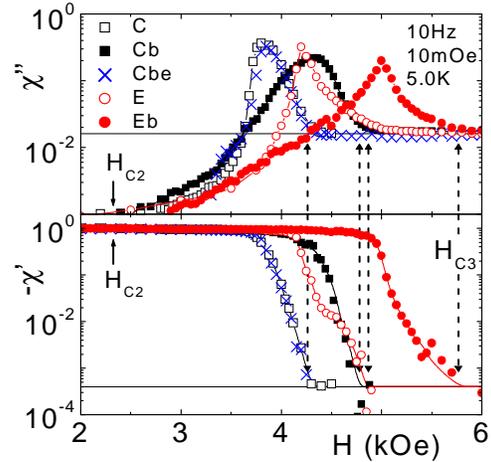}
\caption{Field dependencies of the linear AC-susceptibility on a logarithmic scale of all cylinders measured at 5.0~K,  $10~\mbox{Hz}$ and $H_{AC}= 10~\mbox{mOe}$. $H_{c3}$ is defined by the onset of the screening, i.e., of a finite $-\chi'$ above the noise level. $H_{c3}$ can also be determined from the onset of an excess absorption, see dashed arrows.}
\label{Hc3}
\end{figure}

In order to characterize the superconducting behaviour of the surfaces of all samples, we show in Fig.~\ref{Hc3} the screening and the loss components of the linear AC susceptibilities at 5.0~K: the transition always happens above the upper critical field $H_{c2}$.
Approaching the transition from the normal conducting state, i.e. from high fields, superconductivity appears first at $H_{c3}$ (see Fig.~\ref{Hc3}).
Using the commonly accepted criterion~\cite{hopkinsfinnemore}, originally proposed by Rollins and Silcox~\cite{rollinssilcox}, the nucleation field $H_{c3}$ is defined by the onset of screening, i.e., of a finite $-\chi'$ above the noise level (see caption of Fig.~\ref{Hc3}) .

\begin{figure}[htb]
\centering
\includegraphics*[width=65mm]{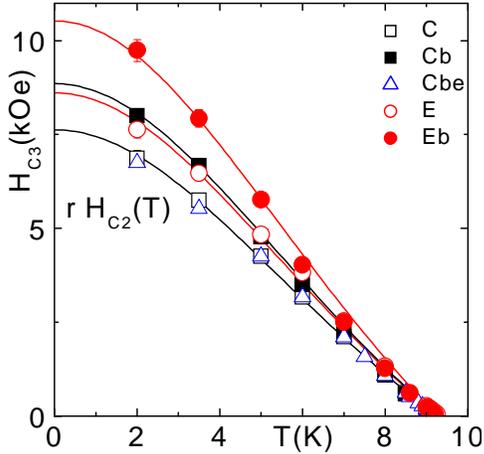}
\caption{ Temperature variations of $H_{c3}$ of all the cylinders as determined in Fig.~\ref{Hc3}. The data are fitted to the function $r H_ {c2}$, where $r$ is a constant depending on the surface preparation.
The different values of $r$ are listed in Table~\ref{r}.}
\label{Hc3T}
\end{figure}

The temperature dependence of $H_{c3}$ of all cylinders is shown in Fig.~\ref{Hc3T}. The data are well described by $H_{c3}(T)=r H_{c2}(T)$, where the ratio $r$ depends on the surface preparation, see Table~\ref{r}.
EP increases $r$ by about $12\%$.
LTB applied to EP and BCP surfaces increases $r$ by about $20\%$ and $\sim 15\%$, respectively.
As can be observed in Figs.~\ref{Hc3} and~\ref{Hc3T}, after removal of $10~\mu\mbox{m}$ from a Cb cylinder (cyl. Cbe) the effect of LTB disappears. These results will be discussed in more detail later (see Discussion).

\begin{figure}[htb]
\centering
\includegraphics*[width=65mm]{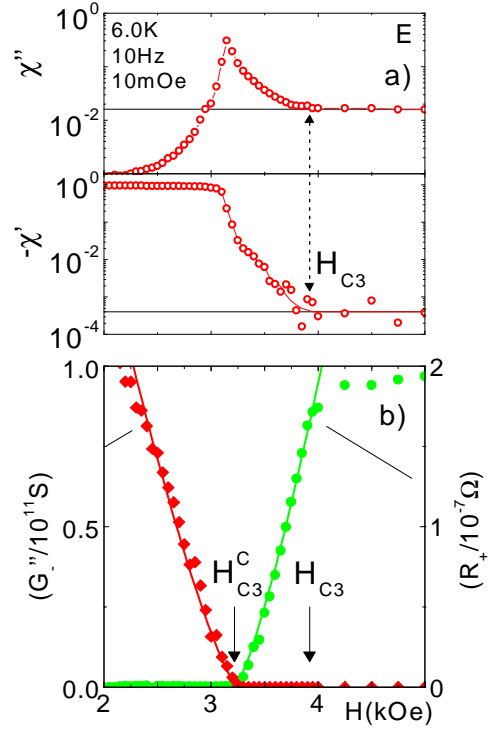}
\caption{ a) Linear AC susceptibility of the E cylinder at 6.0~K and 10~Hz. b) Screening component of the surface conductance $G''$ (diamonds), and the surface resistance $R$ (circles) evaluated from $\chi'-i\chi''$ in a). The singular behaviour of $R_+$ and $G_-''$ near the transition to coherent SSC is described by power laws in 
$\left|1-H/H^{c}_{c3}\right|$ as shown by the solid curves, which at the given low frequency of 10~Hz, unambiguously define the transition field, $H^{c}_{c3}$, from above and below, respectively.}
\label{GR}
\end{figure}

We now examine the complex surface conductance $G~=~2a\sigma$ below $H_{c3}$, where $a$ is the radius of the cylinder and $\sigma$ is the AC conductivity, which has been calculated from the linear AC susceptibility (see also Ref.~\cite{Lars}).
As shown in Fig.~\ref{GR}, between $H_{c3}$ and $H_{c3}^c$  the surface behaviour is still Ohmic:
the resistance $R_+=1/G^{'}(H>H^c_{c3})$ is sharply dropping, but there is only extremely small screening. 
Long-range superconductivity appears only at the lower field $H_{c3}^c$, that we call surface coherent critical field. 
There the screening component of the surface conductance $G''$ rises rapidly, while the surface resistance $R$ vanishes in the limit of low frequencies.
The singular behaviour of $R_+$ and $G_-''$ near the transition to coherent surface superconductivity can be described by power laws in 
$\left|1-H/H^{c}_{c3}\right|$. 
Above the phase transition $R_+\propto (H-H_{c3}^c)^\gamma$, while below $G''_-\propto (H_{c3}^c-H)^\nu$.
For the EP samples (baked and unbaked) we have found $\gamma=\nu=1.3(1)$. This result is consistent with a 2-D model of percolation driven transition to coherent surface superconductivity~\cite{straley}.
For the BCP samples (baked and unbaked) the exponents $\gamma=1.05(10)$ and $\nu=1.4(1)$ indicate a higher dimensionality of the percolating network: still smaller than three but slightly
higher than two~\cite{straley}.
\begin{figure}[htb]
\centering
\includegraphics*[width=65mm]{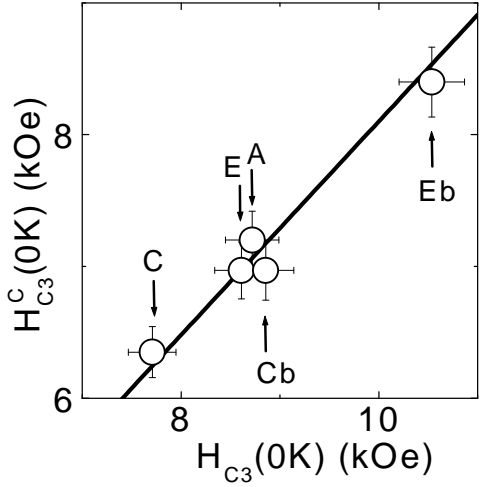}
\caption{Linear relation between $H^{c}_{c3}$ and $H_{c3}$ obtained for the different cylinders. Included is also the result (cyl.~A) for a rougher sample surface, reported  in Ref.~\cite{Lars}.}
\label{Hcc3}
\end{figure}

Figure~\ref{Hcc3} demonstrates, that the surface coherent critical field  is always $H_{c3}^c=0.81(2)H_{c3}$ independent on the electronic structure and surface topology. This interesting result suggests some intrinsic effect behind the formation of the coherent SSC. 

\begin{figure}[htb]
\centering
\includegraphics*[width=65mm]{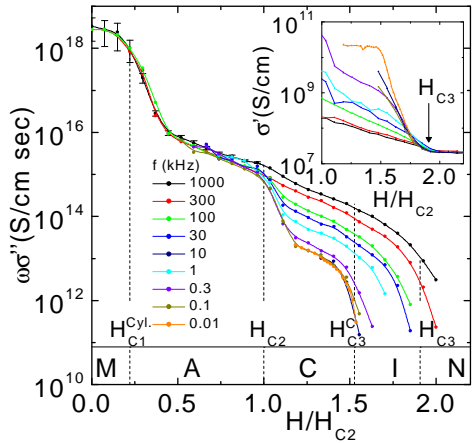}
\caption{Inverse kinetic inductivity $\omega \sigma''$ and conductivity $\sigma'$ (inset) of cylinder C at 4.2~K in magnetic fields up to 2.2 $H_{c2}$, evaluated from linear AC-susceptibilities measured between 10~Hz and 1~MHz. The indicated critical fields separate the superconducting Meissner (M), vortex lattice (A), coherent surface (C), incoherent surface (I), and the normal (N) conducting phases.}
\label{C4e2K}
\end{figure}

Up to now we have discussed only AC susceptibilities and conductivities at 10~Hz.
We have extended our investigations to frequencies up to 1~MHz.
In Fig.~\ref{C4e2K} are depicted the inverse kinetic inductivity $\omega \sigma''$(main frame), which measures the superfluid density, and also the loss component $\sigma '$. 
Both were obtained from the AC susceptibilities measured on the C cylinder at 4.2~K as a function of the applied DC magnetic field.
In the normal conducting region, above $H_{c3}$, $\sigma '(H>H_{c3})=\sigma_n=2.1(3) 10^7~\mbox{S/cm}$ is frequency independent. 
This value is also field independent since it agrees with the one obtained in zero field (see Bulk Superconductivity).
Only below $H_{c3}$,  $\sigma '$ becomes frequency dependent.
From the field dependence of the inverse kinetic inductivity $\omega \sigma''$ it
is possible to distinguish the five phases of the samples obtained by sweeping the DC magnetic field.
The indicated critical fields separate the superconducting Meissner (M), Abrikosov or vortex lattice (A), coherent surface (C), incoherent surface (I), and the normal (N) conducting states.
Below $H_{c2}$, both the vortex response and the Meissner phase are frequency independent.
Interesting to note is that in zero field  the AC penetration depth obtained with large uncertainty $\lambda_{AC}=50(30)~\mbox{nm}$ from $\lambda^{-2}=\mu_0 \omega \sigma''$, is consistent with our DC value $\lambda_{GL}(0)=46(2)~\mbox{nm}$.
The strongest frequency variation is observed close to $H^{c}_{c3}$ indicating a low characteristic frequency, $\omega_c\approx 10^8~\mbox {rad/s}$, of the SSC fluctuations.
\begin{figure}[htb]
\centering
\includegraphics*[width=65mm]{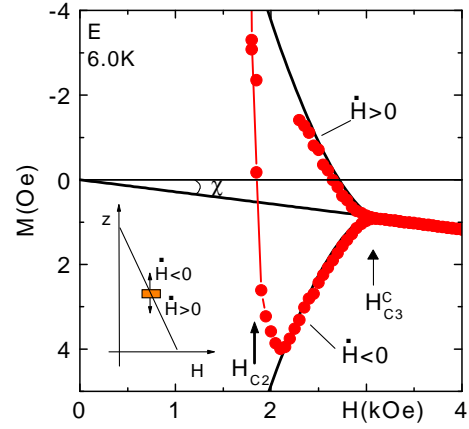}
\caption{Magnetizations of cylinder E between $H_{c2}$ and $H^{c}_{c3}$ at 6.0~K measured along positive and negative field gradients, see inset.}
\label{Ms}
\end{figure}

\begin{figure}[htb]
\centering
\includegraphics*[width=65mm]{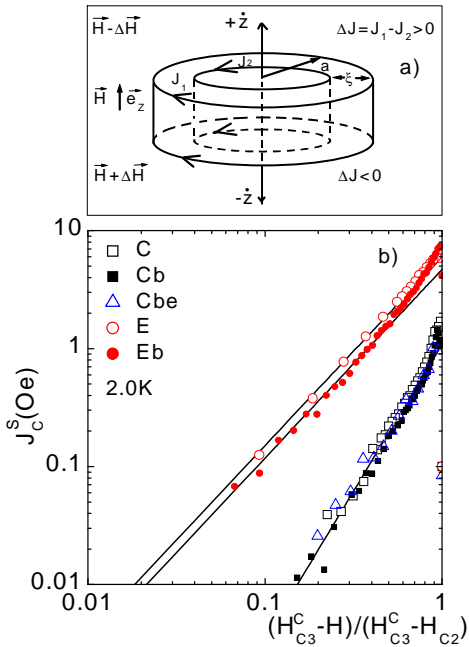}
\caption{a) Sketch of the field-step experiments and the surface supercurrents according to Fink and Barnes model~\cite{finkbarnes}. b) Decay of $J_{c}^s=M-\chi H$ for all cylinders at 2.0~K above $H_{c2}$ as a function of $(H^c_{c3}-H)/(H^c_{c3}-H_{c2})$; note that $1~\mbox{Oe}=10^3/4\pi~\mbox{A/m}$.}
\label{Mc}
\end{figure}

For a clean superconductor (no pinning) an analogy exists between the vortex (A) and the incoherent surface (I) phases: both are Ohmic, incoherent and with zero critical current.
Also the Meissner (M) and the surface coherent (C) phases are analogous since both are coherent critical states supporting supercurrents flowing in a layer of the order of the penetration depth $\lambda$ and in a layer of the order of the coherence length $\xi$, respectively.
To investigate the latter feature, we present evidence for the onset of a surface critical current density at $H^{c}_{c3}$.  
To this end, we moved the sample in a field gradient $\Delta H/\Delta z = 10^{-3}~\mbox{H/cm}$ (see inset Fig.~\ref{Ms}) to measure the response of the magnetization above $H_{c2}$. 
As an example, we show in Fig.~\ref{Ms} the results obtained on the E cylinder at 6.0~K.
According to Lenz' rule, along the positive (negative) field gradient a diamagnetic (paramagnetic) response is found. 
The paramagnetic and the diamagnetic responses are symmetric with respect to the linear background, $\chi H$, which arises from the magnetism of the normal conducting state of the cylinder to be discussed later.   
By subtracting $\chi H$, we obtain the contribution from the induced surface critical current density $J^s_c=M-\chi H$~\cite{lars2}, which we show in Fig.~\ref{Mc} as a function of the reduced field $(H_{c3}^c-H)/(H_{c3}^c-H_{c2})$ at 2.0~K for all samples.
The field dependence of the surface critical current density 
has been  calculated for the first time by Abrikosov~\cite{abrikosov} arriving at a power law in $H_{c3}^c-H$ of the form:
\begin{equation}
J_c^s(H)=\pm J_c^s(H_{c2})\left((H_{c3}^c-H)/(H_{c3}^c-H_{c2})\right)^\alpha,
\end{equation}
with $\alpha =1.5$.
This result was obtained within the GL theory considering one current flowing at the surface (singly connected surface sheath).
For the EP samples we find $\alpha=1.6(1)$ consistent with the Abrikosov calculation, however
the predicted amplitude, $J_c^s(H_{c2})\approx 300~\mbox{Oe}$, is about two orders of magnitudes                                                                                                                                                                                            
higher than the measured values, see Table~\ref{Jc}. 
On the other hand, our results for the EP samples  are in good agreement with the values predicted by the model of Fink and Barnes~\cite{finkbarnes}, who have considered a multiply connected surface sheath with two currents flowing in the opposite direction, illustrated by Fig.~\ref{Mc}~a). 
The resulting  amplitude is given by
\begin{equation}
J_c^s(H_{c2})=\eta H_c \sqrt{(2\lambda/a)} ~F(H/H_{c2})\simeq \eta~ 8~\mbox{Oe},
\end{equation}
where $\eta$ is a factor of the order of 1 and $F(H/H_{c2})$ is tabulated in Ref.~\cite{finkbarnes}.

The $J_c^s(H_{c2})$ of the BCP cylinders are about six times smaller than the $J_c^s(H_{c2})$ of the EP ones.
Moreover, for the BCP samples also the exponent is larger $\alpha=2.5(3)$.

The lower surface current densities $J^s_c$ supported by the BCP surfaces might be an effect of the larger roughness near the surface grain boundaries.
LTB has no significant effect for all cylinders.
\begin{table}[hbt]
\begin{center}
\caption{ Comparison of $J_c^s(H_{c2})$ and of the exponent $\alpha$ obtained from the fits shown in Fig.~\ref{Mc}~b).}
\vskip 0.2truecm
\begin{tabular}{|l|c|c|}
\hline
Sample & $J_c^s(H_{c2})$~(Oe) & $\alpha$ \\ \hline
C, Cb, Cbe    &1.1(2) & 2.5(3)\\ \hline
E & 6.0(3) & 1.6(1) \\ \hline
Eb   & 4.7(3) & 1.6(1) \\ \hline
Ref.~\cite{finkbarnes}  & $\approx 8$ & 1.5 \\ \hline
Ref.~\cite{abrikosov}  &$\approx 300$&1.5 \\ \hline
\end{tabular}
\label{Jc}
\end{center}
\end{table}

\section{DISCUSSION}
While the bulk properties are not affected by the various surface treatments, a rather dramatic change is observed on the surface nucleation fields $H_{c3}$. Table~\ref{r} summarizes the ratios $r=H_{c3}/H_{c2}$ for all samples.
\begin{table}[hbt]
\begin{center}
\caption{Ratios $r=H_{c3}/H_{c2}$ for all samples.}
\vskip 0.2truecm
\begin{tabular}{|l|c|}
\hline
Sample & $r=H_{c3}/H_{c2}$  \\ \hline
GL~\cite{Stjames}    &1.695 \\ \hline
C    &1.86(3) \\ \hline
Cb    &2.16(3) \\ \hline
Cbe    &1.86(3) \\ \hline
E    &2.10(3) \\ \hline
Eb    &2.57(2) \\ \hline
BCS~\cite{hukorenman}    &1.925-5.22 \\ \hline
\end{tabular}
\label{r}
\end{center}
\end{table}

For pure samples the BCS rather than the GL result (see Table~\ref{r}) should apply. 
Then a decrease of the nucleation field $H_{c3}=r H_{c2}$ can only result from surface roughness 
$\overline{\delta r} \gtrsim \xi \approx 50~\mbox{nm}$, which is not observed in the AFM images~\cite{gil}.
Hence, we have to take into account the effects of impurities.

Within the BCS theory, Hu~\cite{hu} predicts a decrease of $r$ by assuming a surface layer with reduced $T_c^s=T_c-\delta T_c$: $r=r_{BCS}(1-2\delta T_c/T_c)$.
This model can be discarded since it implies large variations of the critical temperature at the surface $\delta T_c/T_c\gtrsim 0.15~\mbox{K}$, which are not observed.

For all samples $r$ is larger than $r_{GL}=1.695$.
Within the GL theory the simplest idea is to associate larger $r$'s to an increase of $H^s_{c2}>H_{c2}$, due to a reduction of the coherence length at the surface caused by impurities.
If we suppose mainly oxygen (O) atoms we can determine their
concentration $c_O$ from an expression valid at 4.2~K, $c_O~=~1.475~\cdot~10^{4}(H^s_{c2}(\mbox{Oe})-2760)$~at. $\%$O~\cite{dasgupta}. The results being summarized in Table~\ref{model}, show increasing $c_O$ by EP and LTB.
\begin{table}[hbt]
\begin{center}
\caption{Naive~\cite{dasgupta} and the Shmidt~\cite{shmidt} models.}
\begin{tabular}{|l|c|c|c|c|c|}
\hline
Model & C,Cbe& Cb & E& Eb & Ref.\\ \hline
$c_O$~(at.$\%$O) &0.035&0.106  &0.092 &0.204 & \cite{dasgupta} \\ \hline
dirty $d$(nm)$\geq$ & 2.5& 6.5 & 6&12 &\cite{shmidt} \\ \hline
clean $\ell$(nm)$\leq$ & 436& 169 &192&92&\cite{shmidt}  \\ \hline
\end{tabular}
\label{model}
\end{center}
\end{table}

The model of Fink and Joiner~\cite{finkjoiner} predicts $r>r_{GL}$ for a surface layer 
with an increased $T_c^s=T_c+\delta T_c$ with respect to the one of the bulk.
This model can also be discarded because it predicts an increase of $r$ by approaching $T_c$, which is not observed in Fig.~\ref{Hc3T}, where $r$ is constant.

Another GL-based model assumes that the impurities are contained in a layer of thickness smaller than the coherence length of the bulk $d\leq\xi$~\cite{shmidt}.
This model predicts the following relation:
\begin{equation}
r=1.67\left(1+\left(1-\chi(\xi_0/\ell)\right)\sqrt{1.7}\frac{d}{\xi(T)}\right).
\end{equation}
Here the Gor'kov function~\cite{gorkov}, relates the ratio of the GL parameters of the bulk and the surface, $\chi(\xi_0/\ell)=\kappa/\kappa_s$.
Since $(1-\chi(\xi_0/\ell))\leq 1$ and $d/\xi\leq 1$ the maximum ratio is $r_{max}=3.8$, so that our results are consistent with this model.
Since Eq.(4) embodies two unknowns, i.e. the mean free path $\ell$ and $d$ along with $\xi_0=1.35\xi(0)$~\cite{gorkov}, we can only consider limits.
In the dirty limit $\chi(\xi_0/\ell)_{\ell \rightarrow 0}\simeq 1.33 \ell/\xi_0\rightarrow 0$ we find that $r$ increases if the thickness of the contaminated layer $d$ increases. 
The minimum values of $d$ in the dirty limit are listed in Table~\ref{model}.
In the clean limit $\chi(\xi_0/\ell)_{\ell \rightarrow \infty}\simeq 1-0.884 \xi_0/\ell$, a maximum value for $\ell$ can be determined by assuming $d=\xi$. 
The results listed  in Table~\ref{model}, imply $r$ to increase by a decreasing $\ell$.
Both, this and the naive model are consistent with the idea of O diffusion after baking from the $\sim 5$~nm thin Nb$_2$O$_5$ sheath in a deeper layer~\cite{kneisel}.

\begin{figure}[htb]
\centering
\includegraphics*[width=65mm]{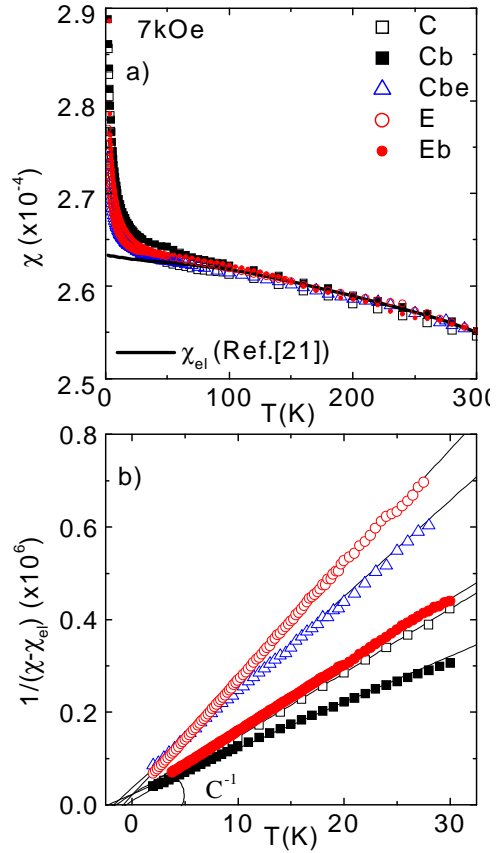}
\caption{ a) Temperature dependence of the susceptibility of all the samples measured at 7~kOe. b) Curie-Weiss contribution to the paramagnetic susceptibility. Solid lines are fits to the Curie-Weiss law $\chi-\chi_{el}=C/T-\theta$. }
\label{fig:curie}
\end{figure}

By measuring the magnetic susceptibility $\chi=M/H_a$ in an external field $H_{a}=7~\mbox{kOe}$ between 2 and $300~ \mbox{K}$, we have also observed an increase of magnetic impurities after baking. 
As shown in Fig.\ref{fig:curie}~a), above $50~\mbox{K}$ our results are in excellent agreement with the bulk values on conducting electrons reported by Hechtfischer~\cite{hechtfischer}, $\chi=\chi_{el}$.
Below $50~\mbox{K}$ an additional contribution is realized, which after removing the contribution due to the conducting electrons reveals for all samples a Curie-Weiss law $\chi-\chi_{el}=C/(T-\theta)$, see Fig.~\ref{fig:curie}~b), where $C$ is the Curie constant and $\theta$ is the Curie-Weiss temperature, see Table~\ref{curie}.
Such behaviour indicates the presence of localized moments due to magnetic impurities.
While  $\theta\simeq -1~\mbox{K}$, we find the Curie constant, being proportional to the concentration of localized magnetic moments, to increase after LTB. 

\begin{table}[hbt]
\begin{center}
\caption{Curie constant $C$ and Curie-Weiss temperatures $\theta$.}
\vskip 0.2truecm
\begin{tabular}{|l|c|c|}
\hline
Sample & $C~(\mu\mbox{K})$ &$\theta$~(K) \\ \hline
C    &72.3(1) & -0.5(2) \\ \hline
Cb    &100.6(7) & -2.2(1) \\ \hline
E    & 40.2(3) & -0.8(2) \\ \hline
Eb    &71.0(1) & -1.5(2) \\ \hline
Cbe    &48.3(4) & -1.7(3)\\ \hline 
\end{tabular}
\label{curie}
\end{center}
\end{table}

O vacancies in the Nb$_2$O$_5$~\cite{cava} sheath would be a good candidate to explain the presence of localized magnetic moments.
Their Curie-Weiss behaviour has been observed in Nb$_2$O$_{5-\delta}$ crystallographic shear structures with $C\simeq 10~\mbox{mK}$ for $\delta\approx 0.17$~\cite{cava}.
However, assuming the shear structure to be present in the $\delta a\simeq 5~\mbox{nm}$ thin Nb$_2$O$_5$ 
layer, one obtains for the Curie constant $C=3\delta a/a \cdot 10~\mbox{mK} \approx 0.12~\mu\mbox{K}$.
Since this is more than two orders of magnitude smaller than the values in Table~\ref{curie},
vacancies of oxygen in Nb$_2$O$_5$ cannot explain the observed magnetic behaviour. 

Unfortunately, the paramagnetism of other impurities, implemented by the BCP and EP processes, like N, C, F, P, S, the hydrogen bonded H$_2$O/C$_x$H$_y$ (OH)$_z$~\cite{halbritter} and some niobium suboxides NbO$_x$ ($x \lesssim 1$), has not yet been investigated.
Perhaps they form clusters with large paramagnetic moments.
After baking, the Curie constant $C$ is found to be increased by about $40-50~\%$.
One possibility would be that during baking additional magnetic moments are released from
external or internal surfaces.
On the other hand, etching away $10~\mu\mbox{m}$ from Cb yields a $50\%$ reduction of $C$, which indeed indicates that
the magnetic moments reside in depths larger than $10~\mu\mbox{m}$ below the surface.
Probably they are localized along the internal surfaces, i.e. grain boundaries or cracks from high temperature annealing.

These localized paramagnetic moments are partially removed after EP, maybe also because of the different chemistry.

\section{SUMMARY AND CONCLUSIONS}

As expected, the bulk properties, i.e. $T_c$, $RRR$, $H_c$, and $H_{c2}$, remain unchanged under the different surface treatments.

\begin{figure}[htb]
\centering
\includegraphics*[width=65mm]{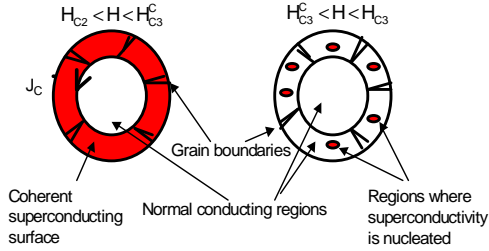}
\caption{Scheme of the surface superconducting phases.}
\label{fig:concl}
\end{figure}

We have shown (see also Ref.~\cite{Lars}) that between the conventional surface
nucleation and the coherent critical field only local superconductivity exists.
Below $H^c_{c3}$ a coherent superconducting phase appears: the SSC screens the cylinder and it supports critical currents (see Fig.~\ref{fig:concl}). 

The power law analyses of the surface conductance and of the surface resistance
of the EP cylinders suggest a 2-dimensional percolation driven transition to coherent surface superconductivity~\cite{Lars, straley}.
For the BCP surfaces indications of a higher dimensionality of the percolating network are found~\cite{straley}.
This may arise from the change of the surface current from the 1-D surface path of thickness $\xi$ in the EP samples to a more 2-D one.
We suspect that this may be related to a fluctuation of weak links along the grain boundaries.
This is consistent with the reduction of the BCP surface critical current density
$J^s_c$ against the EP surfaces and the stronger decay (larger $\alpha$, see Fig.~\ref{Ms} and Table~\ref{Jc}).
LTB has no significant effects on the effective dimensionality of the SSC and on $J_c^s$.

The ratio $H^c_{c3}/H_{c3}$ turns out to be independent on the electronic structure and surface topology, indicating that the nucleation of the surface superconducting coherent state is a hitherto unexplained intrinsic phenomenon.
In contrast, the nucleation field $H_{c3}$ is enhanced by LTB and EP. 
LTB baking increases $H_{c3}$, which we
attributed to a decrease of the normal electron mean free path $\ell$ at the surface due to an increase of impurities.
This is consistent with the diffusion of magnetic impurities into a deeper layer, suggested by the change of the Curie constants.

EP is also increasing $H_{c3}$.
This might be related to an increase of impurities due to the different oxidation process depending on the chemistry of the surface~\cite{halbritter, antoine}. 

\section{ACKNOWLEDGEMENT}

We are indebted to E. A. Knabbe, L. Lilje, P. Schm\"user and B. Steffen (DESY, Hamburg) for providing the Nb cylinders. The work of S.C. was supported by the Deutsche Forschungsgemeinschaft through Grant No. CA 284/1-1.
S.C. and L.v.S. thank the organizers for financial support.

\end{document}